\documentclass[%
 reprint,
superscriptaddress,
preprintnumbers,
 nofootinbib,
 amsmath,amssymb,
 aps,
 prl,
 onecolumn,
]{revtex4-2}

\usepackage{graphicx}
\usepackage{dcolumn}
\usepackage{bm}
\usepackage{physics}
\usepackage{xcolor}
\usepackage{comment}
\usepackage{soul}
\usepackage{upgreek}
\usepackage{tikz}
\usepackage{url}
\usepackage{hyperref}

\newcommand{\mpl}{M_{\mathrm{P}}}
\newcommand{\bx}{\mathbf{x}}
\newcommand{\bp}{\mathbf{p}}

\newcommand{\bk}{\mathbf{k}}
\newcommand{\md}{\mathrm{d}}
\newcommand{\bz}{{\bm \zeta}}

\tikzset{
    cross/.style={path picture={\draw[black]
        (path picture bounding box.south east) -- (path picture bounding box.north west)
        (path picture bounding box.south west) -- (path picture bounding box.north east);}}
}

\begin{document}

\preprint{YITP-25-62}
\preprint{RESCEU-9/25}
\preprint{IPMU25-0018}
\title{Inflationary background renormalization}

\author{Jason Kristiano}
\email{jkristiano@yukawa.kyoto-u.ac.jp}
\affiliation{Center for Gravitational Physics and Quantum Information, Yukawa Institute for Theoretical Physics, Kyoto University, Kyoto 606-8502, Japan}
\affiliation{Research Center for the Early Universe (RESCEU), Graduate School of Science, The University of Tokyo, Tokyo 113-0033, Japan}

\author{Jun'ichi Yokoyama}
\email{junichi.yokoyama@ipmu.jp}
\affiliation{Kavli Institute for the Physics and Mathematics of the Universe (Kavli IPMU), WPI, UTIAS, The University of Tokyo, Kashiwa, Chiba 277-8568, Japan}
\affiliation{Research Center for the Early Universe (RESCEU), Graduate School of Science, The University of Tokyo, Tokyo 113-0033, Japan}
\affiliation{Department of Physics, Graduate School of Science, The University of Tokyo, Tokyo 113-0033, Japan \looseness=-1}
\affiliation{Trans-Scale Quantum Science Institute, The University of Tokyo, Tokyo 113-0033, Japan}


\date{\today}

\begin{abstract}
In cosmic inflation, non-linearities of the curvature perturbation can induce backreaction to the background. To obtain observational predictions at non-linear order on the correct background, one has to redefine the background or introduce background renormalization. We explicitly demonstrate it with a vanishing one-point function of the curvature perturbation as a renormalization condition, so that proper observational predictions can be made even at the nonlinear level. Due to non-linear symmetry of the curvature perturbation, such a procedure induces corrections to the two-point functions, which yield a finite renormalized one-loop correction that depends on the regularization scheme. Cancellation of the divergence is a manifestation of Maldacena's consistency condition. The finite term can be large and highly time-dependent, which indicates evolution outside the horizon.
\end{abstract}

\maketitle
The simplest model of cosmic inflation \cite{Starobinsky:1980te, Sato:1980yn, Guth:1980zm} that is consistent with current observational data \cite{Planck:2018jri, Planck:2019kim} is single-field canonical slow-roll (SR) inflation. It is described by a scalar field $\phi$, called inflaton, with a canonical kinetic term and potential $V(\phi)$ in quasi-de Sitter space. Standard SR inflation generates a nearly scale-invariant adiabatic curvature perturbation that behaves classically as the decaying mode decreases exponentially during inflation, so that the perturbation variable and its conjugate momentum practically commute with each other \cite{Polarski:1995jg}. It is often called superhorizon conservation, although, precisely speaking, the decaying mode exists. In order to realize the SR condition, the shape of the potential is tightly constrained for a finite range of $\phi$.

Nevertheless, superhorizon conservation of curvature perturbation is not a necessary condition for generating a scale-invariant spectrum. Also, adiabaticity in the thermodynamic sense does not imply conservation of perturbation outside the horizon. For example, in the case where the potential $V(\phi)$ is exactly a constant, which is called ultra-slow-roll (USR) inflation \cite{Kinney:1997ne, Inoue:2001zt, Namjoo:2012aa, Martin:2012pe, Motohashi:2014ppa}, the SR condition fails in a specific way, so the perturbation rapidly grows outside the horizon with a scale-invariant spectrum. However, for a phenomenological reason, it cannot explain the observed amplitude of CMB-scale perturbation because an extremely small energy scale of inflation that is lower than the Big Bang Nucleosynthesis energy scale is required \cite{Martin:2012pe}.

Although USR inflation cannot explain the whole 60 e-fold duration of inflation, it may be useful for another purpose. If a USR period with a duration of a few e-folds occurs between SR periods, perturbations which cross the horizon during this period are greatly amplified compared to the earlier one. In general, $V(\phi)$ realizing this \cite{Ivanov:1994pa,Yokoyama:1998pt,Saito:2008em} is called a featured potential. If the power spectrum on small scales reaches $\mathcal{O}(0.01)$, or seven orders greater than the CMB scale $2.1 \times 10^{-9}$, they may collapse to form primordial black holes (PBHs) \cite{Zel:1967, Hawking:1971ei, Carr:1974nx, Carr:2009jm, Carr:2020gox} after they reenter the horizon during the radiation-dominated universe. Despite no observational evidence, PBHs have interesting phenomenological consequences, such as probing Hawking radiation \cite{Hawking:1974rv}, a potential dark matter candidate \cite{Chapline:1975ojl, Garcia-Bellido:1996mdl, Ivanov:1994pa, Yokoyama:1995ex, Yokoyama:1998pt, Afshordi:2003zb, Frampton:2010sw, Belotsky:2014kca, Carr:2016drx, Inomata:2017okj, Espinosa:2017sgp} (reviewed in \cite{Green:2020jor, Carr:2020xqk}), and explaining the origin of gravitational wave events \cite{Sasaki:2016jop, Raidal:2017mfl, Ali-Haimoud:2017rtz, Raidal:2018bbj, Vaskonen:2019jpv, Hall:2020daa}.

Building upon \cite{Kristiano:2021urj, Kristiano:2022zpn}, in our previous papers \cite{Kristiano:2022maq, Kristiano:2023scm, Kristiano:2024vst} (reviewed in \cite{Kristiano:2024ngc}), we computed one-loop corrections to the two-point function or the power spectrum of large-scale perturbations induced by a large amplitude of small-scale perturbations in an inflation model with a featured potential. We obtained an upper bound on the amplitude of small-scale perturbations that must be satisfied so that we can perform renormalization in the perturbative sense. This observation has stimulated lively discussion in the community \cite{Riotto:2023hoz, Riotto:2023gpm, Choudhury:2023vuj, Choudhury:2023jlt, Choudhury:2023rks, Franciolini:2023lgy, Davies:2023hhn, Jackson:2023obv, Mishra:2023lhe, Firouzjahi:2023ahg, Iacconi:2023ggt, Motohashi:2023syh, Tasinato:2023ukp, Tasinato:2023ioq, Maity:2023qzw, Firouzjahi:2023bkt, Firouzjahi:2024psd, Firouzjahi:2023aum, Firouzjahi:2024sce, Firouzjahi:2025gja, Firouzjahi:2025ihn, Sheikhahmadi:2024peu, Fumagalli:2023hpa, Tada:2023rgp,  Cheng:2023ikq, Saburov:2024und, Ballesteros:2024zdp, Ballesteros:2024qqx, Inomata:2024lud, Inomata:2025pqa, Inomata:2025bqw, Fang:2025vhi, Braglia:2025cee, Braglia:2025qrb} because single-field inflation models realizing an appreciable abundance of primordial black holes (PBHs) \cite{Carr:2009jm, Carr:2020gox} are severely constrained. Although we only presented a brief idea on renormalization, we never claimed that there is no counterterm to remove the one-loop correction. Instead, we have presented a condition that must be satisfied by the theory so that counterterms can subtract the loop corrections order by order.

However, higher-order interactions that induce loop corrections to the two-point functions can also generate loop corrections to the one-point function. Then, one might ask whether we do computation on the correct background.   If the comoving curvature perturbation $\zeta$ has a nonvanishing one-point function in nonlinear order, $\langle \zeta \rangle \neq 0$, then the scale factor of the background universe, $a(t)$, would be shifted to $a(t)e^{\langle \zeta(t) \rangle}$, so we would have to redefine the Hubble parameter as $H = \partial_t \log (a e^{\langle \zeta \rangle} )$ under the synchronous condition. In order to avoid changes in background dynamics in non-linear order, we have to make sure that $\langle \zeta \rangle = 0$. In other words, if higher-order interactions induce a nonvanishing one-point function, then we have to renormalize it by imposing $\langle \zeta \rangle = 0$ as a renormalization condition, which is done by introducing linear counterterms. Then, these linear counterterms can induce quadratic counterterms by a nonlinear effect. It was first claimed in \cite{Pimentel:2012tw} within the framework of the effective field theory (EFT) of inflation. Do these quadratic counterterms completely subtract loop corrections to the two-point functions? In other words, does the renormalization condition $\langle \zeta \rangle = 0$ imply $\langle \zeta \zeta \rangle_\mathrm{1-loop} = 0$? This is the question that we would like to answer in this letter.

The action of canonical inflation is given by
\begin{equation}
S = \frac{1}{2} \int \md^4x \sqrt{-g} \left[ \mpl^2 R - (\partial_\mu \phi)^2 - 2 V(\phi) \right], \label{action}
\end{equation}
where $\mpl$ is the reduced Planck scale, and $g = \mathrm{det}~g_{\mu\nu}$, $g_{\mu\nu}$ and $R$ are the metric tensor and its Ricci scalar. Small perturbation from the homogeneous part, $\varphi(t)$, of the inflaton $\phi(\mathbf{x}, t)$ and metric can be expressed as 
\begin{align}
&\phi(\bx,t) = \varphi(t) + \delta \phi(\bx,t), \nonumber\\
&\md s^2(\bx,t) = -N^2(\bx,t)  \mathrm{d}t^2 + \gamma_{ij}(\bx,t) [\mathrm{d}x^i + N^i(\bx,t)  \mathrm{d}t][\mathrm{d}x^j + N^j(\bx,t) \mathrm{d}t],
\end{align}
where $\gamma_{ij}$ is the three-dimensional metric on slices of constant $t$, $N$ is the lapse function, and $N^i$ is the shift vector. We perform analysis in the flat-slicing gauge condition $\gamma_{ij}(\bx,t) = a^2(t) \delta_{ij}$, and then transform to the comoving gauge to obtain the curvature perturbation later. In addition, it is useful to introduce the conformal time $\tau$ with relation $a = -1/(H \tau) \propto e^{H t}$. In this paper, a dot and a prime denote a derivative with respect to $t$ and $\tau$, respectively. Assuming that the first SR parameter is small and negligible in the flat-slicing gauge, namely, taking $N=1$ and  $N^i=0$, we may expand the integrand of action \eqref{action} up to quartic order to yield
\begin{equation}
\mathcal{L} = a^3 \left\lbrace \frac{1}{2} \dot{\varphi}^2 - V(\varphi) + \dot{\varphi} \delta\dot{\phi} - V_1(\varphi) \delta\phi + \frac{1}{2} \left[\delta\dot{\phi}^2 - \frac{1}{a^2} (\partial_i \delta\phi)^2 -V_2(\varphi) \delta\phi^2  \right] - \frac{1}{6} V_3(\varphi) \delta\phi^3 -  \frac{1}{24} V_4(\varphi) \delta\phi^4 \right\rbrace, \label{lag}
\end{equation}
where $V_n \equiv \md^n V/\md \phi^n$. 
Equation of motion of $\delta\phi$ in linear perturbation theory is
\begin{equation}
 \delta\ddot{\phi} +3H\delta\dot{\phi} - \frac{\nabla^2}{a^2} \delta\phi +  V_2(\varphi) \delta\phi = 0 . \label{eom}
\end{equation}
The cubic interaction induces a tadpole diagram to the one-point function
\begin{equation}
\langle \delta\phi(\bx,t) \rangle_\mathrm{bare} = \vcenter{\hbox{
\begin{tikzpicture}
\draw[thick] (0,0) -- (0,0.6);
\draw[thick] (0,-0.2) circle (0.2);
\end{tikzpicture}}}
\neq 0,
\end{equation}
where the horizontal and vertical axes of the schematic diagram represent space and time, respectively.

To subtract the tadpole diagram, one has to perform background renormalization
\begin{equation}
\varphi(t) = \left[ 1 + \frac{1}{2} \delta_Z(t) \right] \varphi_R(t) ~, V(\varphi) = V_R(\varphi) + \delta V(\varphi), \label{bren}
\end{equation}
where $V_R(\varphi)$ is the renormalized potential, $\delta_Z$ and $\delta V(\varphi)$ represent wave function renormalization and counterterm, respectively. The renormalized background field $\varphi_R(t)$ serves as a clock variable in the single-field inflation.
Substituting them into Lagrangian \eqref{lag} and using the linear equation of motion \eqref{eom}, we obtain linear and quadratic counterterms \footnote{In some references \cite{Pimentel:2012tw, Braglia:2025cee, Braglia:2025qrb}, one-point function renormalization is performed within the EFT of inflation framework. In this framework, all possible linear interactions are given by $\mathcal{L}_\mathrm{EFT} \supset - g^{00} \delta M^4 - \delta\Lambda$, where $\delta M^4$ and $\delta\Lambda$ are two independent counterterms. Substituting \eqref{bren} to \eqref{action}, we can map $\delta_Z$ and $\delta V$ to $\delta M^4$ and $\delta \Lambda$.}
\begin{align}
&\mathcal{L}^{(1)}_\mathrm{ct} =  \frac{1}{2} \partial_t \left( a^3 \delta_Z \varphi_R  \delta\dot{\phi} \right) - a^3 \delta V_1(\varphi_R) \delta\phi, \\
&\mathcal{L}^{(2)}_\mathrm{ct} = - \frac{1}{2} a^3 \left[ \delta V_2(\varphi_R) + \frac{1}{2} \delta_Z \varphi_R V_3(\varphi_R) \right] \delta\phi^2.
\end{align}
Note that $\delta V_2(\varphi_R(t))$ can be expressed as $\delta V_2(\varphi_R(t))=\delta \dot{V_1}(\varphi_R(t))/\dot{\varphi}_R(t)$.
Schematically, the linear counterterms induce one-point function
\begin{equation}
\langle \delta\phi(\bx,t) \rangle_\mathrm{ct} = \vcenter{\hbox{
\begin{tikzpicture}
\draw[thick] (0,0) -- (0,0.6) ;
\draw[fill=white,cross] (0,0) circle (0.1) ;
\node[below] at (0,0) {$\scriptstyle{\delta V_1}$};
\end{tikzpicture}}}
+
\vcenter{\hbox{
\begin{tikzpicture}
\draw[thick] (1,0) -- (1,0.6) ;
\draw[fill=white,cross] (1,0) circle (0.1);
\node at (1,0.3) {$\cross$} ;
\node[below] at (1,0) {$\scriptstyle{\delta_Z}$};
\end{tikzpicture}}}
.
\end{equation}
A cross symbol $\times$ in the schematic diagram of counterterm $\delta_Z$ means it is a time integral of a total time derivative. By choosing $\delta V_1(\varphi_R(t)) = - V_3(\varphi_R(t)) \langle \delta\phi^2(t) \rangle / 2$, we obtain the renormalized one-point function as
\begin{equation}
\langle \delta\phi(\bx,t) \rangle_\mathrm{ren} = \langle \delta\phi(\bx,t) \rangle_\mathrm{bare} + \langle \delta\phi(\bx,t) \rangle_\mathrm{ct} =
\vcenter{\hbox{
\begin{tikzpicture}
\draw[thick] (1,0) -- (1,0.6) ;
\draw[fill=white,cross] (1,0) circle (0.1);
\node at (1,0.3) {$\cross$} ;
\node[below] at (1,0) {$\scriptstyle{\delta_Z}$};
\end{tikzpicture}}}
= - \frac{1}{2} \delta_Z(t) \varphi_R(t) .
\end{equation}
Definition and explicit form of $\langle \delta\phi^2(t) \rangle$ will be given in \eqref{dphi2}. The choice of $\delta_Z(t)$ depends on the renormalization condition of $\langle \delta\phi \rangle$. If we impose $\langle \delta\phi \rangle = 0$, then $\delta_Z(t) = 0$, which is implemented in \cite{Inomata:2025bqw, Fang:2025vhi}. However, as explained in the beginning, the proper renormalization condition is $\langle \zeta \rangle = 0$, which differs from $\langle \delta\phi \rangle = 0$, so that the geometrical quantity or the scale factor remains intact.  Indeed only under this condition we can show that the renormalized background $\varphi_R(t)$ satisfies the desired equation of motion 
\begin{equation}
    \ddot{\varphi}_R(t)+3H\dot{\varphi}_R(t)+V_{R1}(\varphi_R)=0,
\end{equation}
with the original $H$.

Going to two-point functions, the bare one-loop corrections induced by the cubic and quartic interactions are
\begin{equation}
\langle \delta\phi(\bx_1,t) \delta\phi(\bx_2,t) \rangle_\mathrm{bare}  = 
\vcenter{\hbox{
\begin{tikzpicture}
\draw[thick] (-0.4,0) arc (-90:-180:0.5);
\draw[thick] (-0.2,0) circle (0.2);
\draw[thick] (0,0) arc (-90:0:0.5);
\end{tikzpicture}}}
+
\vcenter{\hbox{
\begin{tikzpicture}
\draw[thick] (1,0.5) arc (-180:0:0.5);
\draw[thick] (1.5,-0.2) circle (0.2);
\end{tikzpicture}}}
+
\vcenter{\hbox{
\begin{tikzpicture}
\draw[thick] (0,0.5) arc (-180:0:0.5);
\draw[thick] (0.5,0) -- (0.5,-0.6);
\draw[thick] (0.5,-0.8) circle (0.2);
\end{tikzpicture}}}
,
\end{equation}
where the first and second are one-particle irreducible (1PI) and the third diagram is non-1PI. The linear and quadratic counterterms induce two-point functions
\begin{equation}
\langle \delta\phi(\bx_1,t) \delta\phi(\bx_2,t) \rangle_\mathrm{ct}  = \vcenter{\hbox{
\begin{tikzpicture}
\draw[thick] (0,0.5) arc (-180:0:0.5);
\draw[thick] (0.5,0) -- (0.5,-0.6);
\draw[fill=white,cross] (0.5,-0.7) circle (0.1) ;
\node[below] at (0.5,-0.7) {$\scriptstyle{\delta V_1}$};
\end{tikzpicture}}}
+
\vcenter{\hbox{
\begin{tikzpicture}
\draw[thick] (0,0.5) arc (-180:0:0.5);
\draw[thick] (0.5,0) -- (0.5,-0.6);
\draw[fill=white,cross] (0.5,-0.7) circle (0.1) ;
\node[below] at (0.5,-0.7) {$\scriptstyle{\delta_Z}$};
\node at (.50,-0.3) {$\cross$} ;
\end{tikzpicture}}}
+
\vcenter{\hbox{
\begin{tikzpicture}
\draw[thick] (-2,0) arc (-90:-180:0.5);
\draw[fill=white,cross] (-1.9,0) circle (0.1);
\draw[thick] (-1.8,0) arc (-90:0:0.5);
\node[below] at (-1.9,0) {$\scriptstyle{\delta V_2}$};
\end{tikzpicture}}}
+
\vcenter{\hbox{
\begin{tikzpicture}
\draw[thick] (-2,0) arc (-90:-180:0.5);
\draw[fill=white,cross] (-1.9,0) circle (0.1);
\draw[thick] (-1.8,0) arc (-90:0:0.5);
\node[below] at (-1.9,0) {$\scriptstyle{\delta_Z V_3}$};
\end{tikzpicture}}}
,
\end{equation}
where the first and second diagrams are cross-term between cubic interaction and linear counterterms at second-order perturbation and the third and fourth diagrams are quadratic counterterms at first-order perturbation.
Substituting
\begin{equation}
\vcenter{\hbox{
\begin{tikzpicture}
\draw[thick] (-2,0) arc (-90:-180:0.5);
\draw[fill=white,cross] (-1.9,0) circle (0.1);
\draw[thick] (-1.8,0) arc (-90:0:0.5);
\node[below] at (-1.9,0) {$\scriptstyle{\delta V_2}$};
\end{tikzpicture}}}
=
\vcenter{\hbox{
\begin{tikzpicture}
\draw[thick] (-0.2,0) arc (-90:-180:0.5);
\draw[fill=white,cross] (-0.1,0) circle (0.1);
\draw[thick] (0,0) arc (-90:0:0.5);
\node[below] at (-0.1,0) {$\scriptstyle{V_4 \langle \delta\phi^2 \rangle}$};
\end{tikzpicture}}}
+
\vcenter{\hbox{
\begin{tikzpicture}
\draw[thick] (1.4,0) arc (-90:-180:0.5);
\draw[fill=white,cross] (1.5,0) circle (0.1) ;
\draw[thick] (1.6,0) arc (-90:0:0.5);
\node[below] at (1.6,0) {$\scriptstyle{V_3 \partial_t \langle \delta\phi^2 \rangle}$};
\end{tikzpicture}}}
~~\mathrm{and}~~
\vcenter{\hbox{
\begin{tikzpicture}
\draw[thick] (0,0.5) arc (-180:0:0.5);
\draw[thick] (0.5,0) -- (0.5,-0.6);
\draw[fill=white,cross] (0.5,-0.7) circle (0.1) ;
\node[below] at (0.5,-0.7) {$\scriptstyle{\delta_Z}$};
\node at (.50,-0.3) {$\cross$} ;
\end{tikzpicture}}}
+
\vcenter{\hbox{
\begin{tikzpicture}
\draw[thick] (-2,0) arc (-90:-180:0.5);
\draw[fill=white,cross] (-1.9,0) circle (0.1);
\draw[thick] (-1.8,0) arc (-90:0:0.5);
\node[below] at (-1.9,0) {$\scriptstyle{\delta_Z V_3}$};
\end{tikzpicture}}}
= 0,
\end{equation}
we obtain
\begin{equation}
\langle \delta\phi(\bx_1,t) \delta\phi(\bx_2,t) \rangle_\mathrm{ren} =  \langle \delta\phi(\bx_1,t) \delta\phi(\bx_2,t) \rangle_\mathrm{bare} + \langle \delta\phi(\bx_1,t) \delta\phi(\bx_2,t) \rangle_\mathrm{ct} =  
\vcenter{\hbox{
\begin{tikzpicture}
\draw[thick] (-0.4,0) arc (-90:-180:0.5);
\draw[thick] (-0.2,0) circle (0.2);
\draw[thick] (0,0) arc (-90:0:0.5);
\end{tikzpicture}}}
+
\vcenter{\hbox{
\begin{tikzpicture}
\draw[thick] (1.4,0) arc (-90:-180:0.5);
\draw[fill=white,cross] (1.5,0) circle (0.1) ;
\draw[thick] (1.6,0) arc (-90:0:0.5);
\node[below] at (1.6,0) {$\scriptstyle{V_3 \partial_t \langle \delta\phi^2 \rangle}$};
\end{tikzpicture}}}
,
\end{equation}
that is independent of counterterm $\delta_Z(t)$.

Moreover, we are interested in the curvature perturbation. It can be obtained by performing a gauge transformation from flat-slicing to comoving gauge. Defining $\bz = -H \delta\phi/ \dot{\varphi}_R$, the non-linear gauge transformation reads $\zeta = \bz + f(\bz)$, where $f(\bz) = \sum_{n=2}^\infty f^{(n)}(\bz)$. The quadratic order of $f(\bz)$ is given by \cite{Maldacena:2002vr}
\begin{equation}
f^{(2)}(\bz) = \frac{\eta}{4} \bz^2 + \frac{1}{H} \dot{\bz} \bz + \frac{1}{4(a H)^2} [-(\partial_i \bz)^2 + \partial^{-2}\partial_i \partial_j (\partial_i \bz \partial_j \bz)] + \mathcal{O}(\epsilon),
\end{equation}
where $\epsilon\equiv -\dot{H}/H^2$ and $\eta\equiv \dot{\epsilon}/(\epsilon H)$ are the first and second SR parameters, respectively, defined with respect to the renormalized background. Then, the renormalized expectation value of the gauge-transformed one-point function reads $\langle \zeta \rangle_\mathrm{ren} = \langle \bz \rangle_\mathrm{ren} + \langle f^{(2)}(\bz) \rangle$. Imposing a renormalization condition $\langle \zeta \rangle_\mathrm{ren} = 0$ means $\langle \bz \rangle_\mathrm{ren} = -\langle f^{(2)}(\bz) \rangle $, which corresponds to $\langle \delta\phi \rangle_\mathrm{ren} = \dot{\varphi}_R \langle f^{(2)}(\bz) \rangle / H$. Defining $\dot{\varphi}_R \tilde{\delta}_Z/H = \delta_Z \varphi_R/2$, we obtain $\tilde{\delta}_Z = \langle f^{(2)}(\bz) \rangle$. Therefore, we have fixed the two counterterms $\delta V(t)$ and $\delta_Z(t)$, based on the renormalization condition $\langle \zeta \rangle_\mathrm{ren} = 0$.

Before deriving the explicit form of counterterms, we have to evaluate the following quantity
\begin{equation}
\langle \delta\phi^2(\tau) \rangle = \frac{1}{a^2(\tau)} \langle v^2(\tau) \rangle = \int \frac{\md k}{k} \frac{1}{a^2(\tau)} \Delta_v^2(k, \tau), \label{dphi2}
\end{equation}
where $\Delta_v^2(k, \tau) \equiv k^3 \abs{v_k(\tau)}^2 / (2 \pi^2)$ is power spectrum of the Mukhanov-Sasaki variable $v_k(\tau)$. It satisfies the Mukhanov-Sasaki equation
\begin{equation}
v_k''+ \left( k^2 - \frac{z''}{z} \right) v_k = 0,
\end{equation}
where $z \equiv a\sqrt{2\epsilon}$. With WKB approximation, ultraviolet (UV) limit of the Mukhanov-Sasaki variable is
\begin{equation}
\lim_{k \rightarrow \infty} \Delta_v^2(k,\tau) = \frac{k^2}{4\pi^2} + \frac{1}{8 \pi^2} \frac{z''(\tau)}{z(\tau)}, \label{msuv}
\end{equation}
which is well-known in the literatures of adiabatic regularization \cite{Pla:2024xsv}.

To derive the explicit form of $\langle \delta\phi^2(\tau) \rangle$, we have to specify the regularization scheme. With cutoff regularization, the explicit form is
\begin{equation}
\langle \delta\phi^2(\tau) \rangle = \int^{\Lambda a(\tau)}_{k_\mathrm{IR}} \frac{\md k}{k} \frac{1}{a^2(\tau)} \Delta_v^2(k,\tau) = \frac{\Lambda^2}{8\pi^2} + \frac{1}{8\pi^2 a^2(\tau)} \frac{z''(\tau)}{z(\tau)} \log \frac{\Lambda a(\tau)}{k_\mathrm{IR}},
\end{equation}
where a physical UV cutoff $\Lambda$ and a comoving IR cutoff $k_\mathrm{IR}$ are imposed. Performing time derivative yields \footnote{A unique choice of UV cutoff for phase space integral of $\abs{\delta\phi_k(\tau_1)}^2$ is $\Lambda a(\tau_1)$ and $\partial_{\tau_1} \langle \delta\phi^2(\tau_1) \rangle$ is obtained simply by taking time derivative of it. Given $\abs{\delta\phi_k(\tau_1)}^2$, we can obtain $\partial_{\tau_1} ( \abs{\delta\phi_k(\tau_1)}^2)$ simply by taking a derivative of $\abs{\delta\phi_k(\tau_1)}^2$ with respect to $\tau_1$. Alternatively, despite being more complicated, given the same quantity, we can obtain $\partial_{\tau_1} ( \abs{\delta\phi_k(\tau_1)}^2)$ by taking a derivative of $\abs{\delta\phi_k(\tau_1)}^2$ with respect to $k$ as expressed in \eqref{maldacena}. Just because it can be expressed alternatively as \eqref{maldacena} does not imply that the UV cutoff depends on an auxiliary time $\tau_2$. This is the point where we disagree with cutoff regularization of the quadratic counterterm in \cite{Inomata:2025bqw, Fang:2025vhi}. The quadratic counterterm is induced by linear counterterm, which is evaluated at first-order perturbation that only has one bulk time integral that makes it a unique time variable.}
\begin{equation}
\partial_\tau \langle \delta\phi^2(\tau) \rangle = \frac{1}{8\pi^2} \left[ \frac{1}{a^2(\tau)} \frac{z''(\tau)}{z(\tau)} \right]' \log \frac{\Lambda a(\tau)}{k_\mathrm{IR}} + \frac{H}{8\pi^2 a(\tau)} \frac{z''(\tau)}{z(\tau)}.
\end{equation}

Another well-studied regularization scheme is the dimensional regularization (dim-reg). In this scheme, the explicit form is \footnote{In $3+\delta$-dimension, $z = a^{1+\delta}\sqrt{2\epsilon}$, so $z''/z$ is a function of $\delta$. Because of that, we have to expand $(z''/z)_\delta = (z''/z)_0 + \delta \partial_\delta (z''/z)_0$ to obtain the finite part. We have to do the same for $(z''/z)$ in \eqref{intdim}, so after subtraction, the renormalized two-point functions will be the same as \eqref{dimregren}.}
\begin{align}
\langle \delta\phi^2(\tau) \rangle &= \int^{\infty}_{k_\mathrm{IR}}  \frac{\md k}{k} \left( \frac{k}{\mu} \right)^\delta \frac{1}{a^{2+\delta}(\tau)} \Delta_v^2(k,\tau) = - \frac{1}{\delta} \left[ \frac{k_\mathrm{IR}}{\mu a(\tau)} \right]^\delta \frac{1}{8\pi^2 a^2(\tau)} \frac{z''(\tau)}{z(\tau)} \nonumber\\
&=  \frac{1}{8\pi^2 a^2(\tau)} \frac{z''(\tau)}{z(\tau)} \left[ -\frac{1}{\delta} - \log \frac{k_\mathrm{IR}}{\mu a(\tau)} \right],
\end{align}
where $\delta$ is a small correction to the dimension and $\mu$ is a renormalization scale. As expected from Minkowski space QFT, dim-reg cannot capture polynomial divergence $\Lambda^2$. When we encounter an integral of the form $\int \md^3k ~f(k)$, we have to expand $f(k)$ asymptotically as $k \rightarrow \infty$ to obtain a term proportional to $1/k^3$ because dim-reg regulates only a log-divergent integral. In the dim-reg context, we take $\int \md^3k ~k^\alpha = 0$ for the real value of $\alpha \neq -3$. This procedure is explained in detail in \cite{Ballesteros:2024qqx} with an emphasis on computing the finite parts. Performing the time derivative yields
\begin{equation}
\partial_\tau \langle \delta\phi^2(\tau) \rangle = \frac{1}{8\pi^2} \left[ \frac{1}{a^2(\tau)} \frac{z''(\tau)}{z(\tau)} \right]' \left[ - \frac{1}{\delta} + \log \frac{\mu a(\tau)}{k_\mathrm{IR}} \right] + \frac{H}{8\pi^2 a(\tau)} \frac{z''(\tau)}{z(\tau)}.
\end{equation}

In dim-reg, the explicit form of the counterterm and cubic-exchange diagram of the two-point functions read
\begin{align}
\vcenter{\hbox{
\begin{tikzpicture}
\draw[thick] (1.4,0) arc (-90:-180:0.5);
\draw[fill=white,cross] (1.5,0) circle (0.1) ;
\draw[thick] (1.6,0) arc (-90:0:0.5);
\node[below] at (1.6,0) {$\scriptstyle{V_3 \partial_t \langle \delta\phi^2 \rangle}$};
\end{tikzpicture}}} = - \int_{-\infty}^{\tau_0} \md\tau_1 a^{4+\delta}(\tau_1) (2 \epsilon(\tau_1))^{3/2} V_3(\tau_1) 2 \mathrm{Im}\left[ \bz_p(\tau_0) \bz_p^{*}(\tau_1) \right] \mathrm{Re}\left[ \bz_p(\tau_0) \bz_p^{*}(\tau_1) \right] \frac{\partial_{\tau_1} \langle \delta\phi^2(\tau_1) \rangle}{2 \epsilon(\tau_1) a(\tau_1) H} , \label{dimct}
\end{align}
\begin{align}
\vcenter{\hbox{
\begin{tikzpicture}
\draw[thick] (-0.4,0) arc (-90:-180:0.5);
\draw[thick] (-0.2,0) circle (0.2);
\draw[thick] (0,0) arc (-90:0:0.5);
\end{tikzpicture}}} = & ~4 \int_{-\infty}^{\tau_0} \md\tau_1 a^{4+\delta}(\tau_1) (2 \epsilon(\tau_1))^{3/2} V_3(\tau_1) \mathrm{Im}\left[ \bz_p(\tau_0) \bz_p^{*}(\tau_1) \right] \nonumber\\
& \times \int_{-\infty}^{\tau_1} \md \tau_2 a^{4+\delta}(\tau_2) (2 \epsilon(\tau_2))^{3/2}  V_3(\tau_2) \int_{k_\mathrm{IR}}^\infty \frac{\md^{3+\delta}k}{(2\pi)^{3+\delta}} \mathrm{Im}\left[ \bz_p(\tau_0) \bz_p^*(\tau_2) \bz_k(\tau_1) \bz_q(\tau_1) \bz_k^*(\tau_2) \bz_q^*(\tau_2) \right], \label{dimex}
\end{align}
where $q=\abs{\bk-\bp} $. For $p \ll k$, the sum becomes
\begin{align}
& \vcenter{\hbox{
\begin{tikzpicture}
\draw[thick] (1.4,0) arc (-90:-180:0.5);
\draw[fill=white,cross] (1.5,0) circle (0.1) ;
\draw[thick] (1.6,0) arc (-90:0:0.5);
\node[below] at (1.6,0) {$\scriptstyle{V_3 \partial_t \langle \delta\phi^2 \rangle}$};
\end{tikzpicture}}} +
\vcenter{\hbox{
\begin{tikzpicture}
\draw[thick] (-0.4,0) arc (-90:-180:0.5);
\draw[thick] (-0.2,0) circle (0.2);
\draw[thick] (0,0) arc (-90:0:0.5);
\end{tikzpicture}}} = 2 \int_{-\infty}^{\tau_0} \md\tau_1 a^{4+\delta}(\tau_1) (2 \epsilon(\tau_1))^{3/2} V_3(\tau_1) \mathrm{Im}\left[ \bz_p(\tau_0) \bz_p^{*}(\tau_1) \right] \abs{\bz_p(\tau_0)}^2 \nonumber\\
& \times \left\lbrace -\frac{\partial_{\tau_1} \langle \delta\phi^2(\tau_1) \rangle}{2 \epsilon(\tau_1) a(\tau_1) H} + 2 \int_{-\infty}^{\tau_1} \md \tau_2 a^{4+\delta}(\tau_2) (2 \epsilon(\tau_2))^{3/2}  V_3(\tau_2) \int_{k_\mathrm{IR}}^\infty \frac{\md^{3+\delta}k}{(2\pi)^{3+\delta}} \mathrm{Im}\left[ \bz_k^2(\tau_1) \bz_k^{*2}(\tau_2) \right] \right\rbrace. \label{dimregsum}
\end{align}
Deep inside the horizon, $k \rightarrow \infty$, the mode function approaches
\begin{equation}
\bz_k(\tau) \rightarrow \frac{e^{-i k \tau}}{2a^{1+ \frac{1}{2}\delta}(\tau)\sqrt{\epsilon(\tau) k}},
\end{equation}
we can evaluate the time integral as \footnote{One may evaluate the wavenumber integral before the time integral and obtain the exactly same result. When evaluating the time integral, we assume that the time-dependent quantities are constant within time scale $1/k$, which makes sense when $k\rightarrow\infty$. Also, every time integral with lower bound $-\infty$ implicitly includes the $i\varepsilon$ prescription as $-\infty(1-i\varepsilon)$.}
\begin{align}
&2 \int_{k_\mathrm{IR}}^\infty \md k \left[ \frac{k}{\mu a(\tau_1)} \right]^\delta \int_{-\infty}^0 \md(\Delta\tau) \frac{ a^4(\tau_1+\Delta\tau) V_3(\tau_1+\Delta\tau) (2\epsilon(\tau_1+\Delta\tau))^{3/2} }{32\pi^2 a^2(\tau_1+\Delta\tau)\epsilon(\tau_1+\Delta\tau) a^2(\tau_1)\epsilon(\tau_1)} \mathrm{Im} e^{2i k \Delta\tau} \nonumber\\
=& - 2 \int_{k_\mathrm{IR}}^\infty \frac{\md k}{2k} \left[ \frac{k}{\mu a(\tau_1)} \right]^\delta \frac{ a^4(\tau_1) V_3(\tau_1) (2\epsilon(\tau_1))^{3/2} }{32\pi^2 a^4(\tau_1)\epsilon^2(\tau_1) } = \frac{ a^4(\tau_1) V_3(\tau_1) (2\epsilon(\tau_1))^{3/2} }{32\pi^2 a^4(\tau_1) \epsilon^2(\tau_1) } \left[ \frac{1}{\delta} + \log \frac{k_\mathrm{IR}}{\mu a(\tau_1)} \right]. \label{intdim}
\end{align}
With relation
\begin{equation}
V_3(\tau) = - \frac{H}{a(\tau)\sqrt{2\epsilon(\tau)}} \left[ \frac{1}{a^2(\tau)} \frac{z''(\tau)}{z(\tau)} \right]',
\end{equation}
we can subtract this cubic-exchange with the counterterm diagram to obtain a UV finite term contribution to the two-point functions
\begin{equation}
\left\langle \zeta_\bp(\tau_0)\zeta_{-\bp}(\tau_0) \right\rangle_\mathrm{ren} = -2  \abs{\bz_p(\tau_0)}^2 \int_{-\infty}^{\tau_0} \md\tau_1 a^4(\tau_1) (2 \epsilon(\tau_1))^{3/2} V_3(\tau_1) \mathrm{Im}\left[ \bz_p(\tau_0) \bz_p^{*}(\tau_1) \right] \left[ \frac{1}{16 \pi^2 \epsilon(\tau_1) a^2(\tau_1)} \frac{z''(\tau_1)}{z(\tau_1)} \right]. \label{dimregren}
\end{equation}
The counterterms are used to remove the divergent component of the cubic-exchange diagram. A priori, one should not assume that the finite term is small or a constant. It can be large and highly time-dependent. Note that the counterterm is completely fixed by the one-point function renormalization condition, even up to the finite part. Thus, there is no ambiguity on the finite part of the counterterm, so we cannot absorb this finite part of the one-loop two-point functions to the counterterms associated with background renormalization.

Given the mode function $\delta\phi_k(\tau)$, what is $\partial_\tau \langle \delta\phi^2(\tau) \rangle$? Of course, we can derive it by computing $\langle \delta\phi^2(\tau) \rangle$ and taking the derivative with respect to $\tau$. However, with great mathematical curiosity, can we take a derivative with respect to $k$ and obtain $\partial_\tau \langle \delta\phi^2(\tau) \rangle$? The answer is manifested in Maldacena's consistency relation \cite{Maldacena:2002vr}
\begin{align}
- \frac{\md\log \Delta_\bz^2(k,\tau_1)}{\md\log k} \abs{\bz _k(\tau_1)}^2 = &- 2 \int_{-\infty}^{\tau_1} \md \tau_2 a^{4+\delta}(\tau_2) V_3(\tau_2) (2\epsilon(\tau_2))^{3/2} \mathrm{Im}[ \bz_k^2(\tau_1) \bz_k^{*2}(\tau_2) ] \nonumber\\
&+ \eta(\tau_1) \abs{\bz_k(\tau_1)}^2 + \frac{2}{a(\tau_1)H} \mathrm{Re}[ \bz_k(\tau_1) \bz_k'^*(\tau_1)], \label{maldacena}
\end{align}
where $\Delta_\bz^2(k, \tau)$ is the power spectrum of $\bz_k(\tau)$ at $3+\delta$-dimension\footnote{In case $V(\tau)$ realizes sharp transition of $\eta(\tau)$ from USR to SR period, this consistency relation is explicitly shown in \cite{Kristiano:2023scm}.}. The second line can be recast to $\partial_{\tau_1}  \left( \abs{\delta\phi_k(\tau_1)}^2 \right)$. Taking the wavenumber integral reads
\begin{equation}
\int \frac{\md^{3+\delta} k}{(2\pi)^{3+\delta}} \left\lbrace   2 \int_{-\infty}^{\tau_1} \md \tau_2 a^{4+\delta}(\tau_2) V_3(\tau_2) (2\epsilon(\tau_2))^{3/2} \mathrm{Im}[ \bz_k^2(\tau_1) \bz_k^{*2}(\tau_2) ] - \frac{\partial_{\tau_1}  \left( \abs{\delta\phi_k(\tau_1)}^2 \right) }{2 \epsilon(\tau_1) a(\tau_1) H} =  \frac{\md\log \Delta_\bz^2(k,\tau_1)}{\md\log k} \abs{\bz _k(\tau_1)}^2 \right\rbrace, \label{intcr}
\end{equation}
we can read that the left-hand side is exactly the second line of \eqref{dimregsum}. The right-hand side gives a finite term
\begin{equation}
\int_{k_\mathrm{IR}}^\infty \md k ~\frac{\md}{\md k}\left\lbrace \left[ \frac{k}{\mu a(\tau_1)} \right]^\delta \frac{\Delta_v^2(k,\tau_1)}{2\epsilon(\tau_1) a^2(\tau_1)} \right\rbrace = \frac{1}{16\epsilon(\tau_1) a^2(\tau_1)} \frac{z''(\tau_1)}{z(\tau_1)} \int_{k_\mathrm{IR}}^\infty \frac{\md k}{k} \delta k^\delta = - \frac{1}{16\epsilon(\tau_1) a^2(\tau_1)} \frac{z''(\tau_1)}{z(\tau_1)},
\end{equation}
which exactly coincides with \eqref{dimregren}. However, it should be noted that we previously regularized the cubic exchange and counterterm diagrams one by one using a UV approximation of the Mukhanov-Sasaki variable \eqref{msuv}. In this alternative approach, the subtraction \eqref{intcr} holds exactly without relying on the UV approximation \eqref{msuv}\footnote{Of course, if one substitutes \eqref{msuv} to \eqref{maldacena} and \eqref{intcr}, one finds the left- and right-hand sides are equal.}. Finally, we evaluated the right-hand side of \eqref{intcr} with dim-reg by extracting the log-divergent integral with the UV approximation \eqref{msuv} to obtain the UV finite term.

In the cutoff regularization scheme, for second-order perturbation theory, following \cite{Senatore:2009cf}, one has to regularize both the time and wavenumber integral
\begin{align}
\vcenter{\hbox{
\begin{tikzpicture}
\draw[thick] (-0.4,0) arc (-90:-180:0.5);
\draw[thick] (-0.2,0) circle (0.2);
\draw[thick] (0,0) arc (-90:0:0.5);
\end{tikzpicture}}} = & ~4 \int_{-\infty}^{\tau_0} \md\tau_1 a^4(\tau_1) (2 \epsilon(\tau_1))^{3/2} V_3(\tau_1) \mathrm{Im}\left[ \bz_p(\tau_0) \bz_p^{*}(\tau_1) \right] \nonumber\\
& \times \int_{-\infty}^{\tau_1\left(1+\frac{H}{\Lambda} \right)} \md \tau_2 a^4(\tau_2) (2 \epsilon(\tau_2))^{3/2}  V_3(\tau_2) \int_{k_\mathrm{IR}}^{\Lambda a(\tau_2)} \frac{\md^3 k}{(2\pi)^3} \mathrm{Im}\left[ \bz_p(\tau_0) \bz_p^*(\tau_2) \bz_k(\tau_1) \bz_q(\tau_1) \bz_k^*(\tau_2) \bz_q^*(\tau_2) \right].
\end{align}
For $p \ll k$, the integral becomes \footnote{In contrast to \eqref{intdim}, we must compute the wavenumber integral before the time integral. In this case, the time integral diverges if its upper bound is zero}
\begin{align}
& \int_{-\infty}^{\tau_1\frac{H}{\Lambda} } \md (\Delta\tau) \frac{ a^4(\tau_1+\Delta\tau) (2 \epsilon(\tau_1+\Delta\tau)^{3/2}  V_3(\tau_1+\Delta\tau)}{32\pi^2 a^2(\tau_1+\Delta\tau) \epsilon(\tau_1+\Delta\tau) a^2(\tau_1) \epsilon(\tau_1)} \int_{k_\mathrm{IR}}^{\Lambda a(\tau_1+\Delta\tau)} \md k ~\mathrm{Im} e^{2i k \Delta\tau} \nonumber\\
=& \int_{-\infty}^{\tau_1\frac{H}{\Lambda} } \md (\Delta\tau) \frac{ a^4(\tau_1+\Delta\tau) (2 \epsilon(\tau_1+\Delta\tau)^{3/2}  V_3(\tau_1+\Delta\tau)}{32\pi^2 a^2(\tau_1+\Delta\tau) \epsilon(\tau_1+\Delta\tau) a^2(\tau_1) \epsilon(\tau_1)} \frac{-1}{2\Delta\tau} \left\lbrace \cos \left[ \frac{\Lambda \Delta\tau}{H (\tau_1+\Delta\tau)} \right] - \cos \left[ k_\mathrm{IR} \Delta\tau \right] \right\rbrace \nonumber\\
\simeq &~\frac{a^4(\tau_1) V_3(\tau_1) (2\epsilon(\tau_1))^{3/2}}{64 \pi^2 a^4(\tau_1) \epsilon^2(\tau_1)} \left[ \gamma_\mathrm{E} + \log \frac{k_\mathrm{IR}}{\Lambda a(\tau_1)} \right],
\end{align}
where $\gamma_\mathrm{E}$ is the Euler–Mascheroni constant. Subtracting it with counterterm yields
\begin{align}
\left\langle \zeta_\bp(\tau_0)\zeta_{-\bp}(\tau_0) \right\rangle_\mathrm{ren} = \abs{\bz_p(\tau_0)}^2 \int_{-\infty}^{\tau_0} & \md\tau_1 a^4(\tau_1) (2 \epsilon(\tau_1))^{3/2} V_3(\tau_1) \mathrm{Im}\left[ \bz_p(\tau_0) \bz_p^{*}(\tau_1) \right] \nonumber\\
&\times \left[ - \frac{1}{8\pi^2 \epsilon(\tau_1) a^2(\tau_1)} \frac{z''(\tau_1)}{z(\tau_1)} + \frac{\gamma_\mathrm{E}}{16 \pi^2\epsilon^2(\tau_1)} (2\epsilon(\tau_1))^{3/2}V_3(\tau_1) \right].
\end{align}

It is not surprising that a different regularization scheme yields a different UV finite term, which also happens in flat space QFT. However, in flat space QFT, it is not relevant because it can be absorbed by the finite part of quadratic counterterms depending on the renormalization condition of the two-point functions themselves. Here, we cannot do that because the quadratic counterterms are induced from linear counterterms that are completely fixed by the renormalization condition of the one-point function. Thus, we need counterterms beyond background renormalization to completely remove the UV finite term of one-loop two-point functions with a renormalization condition on the two-point functions themselves. These are quadratic counterterms that are not induced from the linear one, for example, the speed of sound \cite{Iacconi:2023ggt} and higher-order derivatives \cite{Braglia:2025cee, Braglia:2025qrb} counterterms. After further renormalization, the two-point functions can be compared with observations. We expect that our result can be generalized to higher-point functions in which their loop corrections are not completely subtracted by background renormalization.

In this letter, we have performed calculations in flat-slicing gauge and obtained the curvature perturbation by gauge transformation to comoving gauge. In our companion paper \cite{Kristiano}, we will perform computations directly in the comoving gauge and show that the same results are obtained. We will also show the relationship with EFT of the non-attractor inflation formalism \cite{Akhshik:2015nfa} that is used in \cite{Firouzjahi:2023aum, Firouzjahi:2023bkt, Firouzjahi:2024psd, Firouzjahi:2024sce, Firouzjahi:2025gja, Firouzjahi:2025ihn}. Moreover, we will derive renormalized two-point functions to a general scale by summing \eqref{dimct} and \eqref{dimex} without assuming $p \ll k$, for a specific inflationary potential that realizes a sharp or smooth transition of the second SR parameter. A more challenging direction is non-perturbative background renormalization in light of recent works on non-perturbative wavefunction \cite{Celoria:2021vjw, Creminelli:2024cge}, lattice simulation \cite{Caravano:2024tlp, Caravano:2024moy}, and non-perturbative Hamiltonian \cite{Firouzjahi:2025ihn, Firouzjahi:2025gja}.

In conclusion, we have done inflationary background renormalization and discussed its implication to one-loop two-point functions of the curvature perturbation. We have shown that the vanishing one-point function of the curvature perturbation can be achieved by introducing two independent linear counterterms. These counterterms induce quadratic counterterms that subtract the divergent part of one-loop two-point functions, resulting in a UV finite term that can be large and highly time-dependent. We have found that such a finite term depends on the regularization scheme and must be taken as it is because the quadratic counterterms are completely fixed by the one-point function renormalization condition.

J.~K. acknowledges the support from JSPS KAKENHI Grant No.~23K25868. J.~Y. is supported by JSPS KAKENHI Grant No.~25K07296. J.~K. is grateful to all the participants of Looping in the Primordial Universe at CERN Department of Theoretical Physics for interesting discussions related to this paper. J.~K. thanks Guilherme Pimentel for his kind hospitality and discussion while staying at Scuola Normale Superiore during the latest stage of this project. J.~K. also acknowledges discussions with some participants of cosmological correlators workshop at National Taiwan University. J.~K. thanks Keisuke Inomata and Cheng-jun Fang for explaining their papers by email exchange.

\bibliographystyle{apsrev4-1}
\bibliography{reference}

\end{document}